**ORIGINAL RESEARCH**  Open Access

# In Search of Newer Targets for Inflammatory Bowel Disease: A Systems and a Network Medicine Approach

Takashi Kitani,[1] Sushma C. Maddipatla,[2] Ramya Madupuri,[2] Christopher Greco,[2] Jonathan Hartmann,[3] James N. Baraniuk,[4] and Sona Vasudevan[2,*]

**Abstract**

**Introduction:** Crohn's disease and ulcerative colitis, both under the umbrella of inflammatory bowel diseases (IBD), involve many distinct molecular processes. The difference in their molecular processes is studied by using the different genes involved in each disease, and it is explored further for drug targeting and drug repurposing.

**Methods:** The initial set of genes was obtained by mining published literature and several curated databases. The identified genes were then subject to Systems and Network analysis to reveal their molecular processes and shed some light on their pathogenesis. Such methodologies have identified newer targets and drugs that can be repurposed.

**Results:** We use a Systems and Network Medicine approach to understand the mechanism of actions of genes involved in IBD. From an initial set of genes mined from literature and curated databases, we used the Multi-Steiner Tree algorithm implemented within the CoVex systems medicine platform to expand each disease module by incorporating candidate genes with significant connections to the disease-related seed genes. Such expanded disease modules will identify a larger set of potential targets and drugs. We used the Closeness Centrality algorithm implemented within CoVex to search for newer targets and repurposable drugs. Through a network medicine approach, we provide a mechanistic view of the diseases and point to newer drugs and targets.

**Conclusion:** We demonstrate that the Systems and Network Medicine approach is a powerful way to understand diseases and understand their mechanisms of action.

## Introduction

Inflammatory bowel disease (IBD) is generally an umbrella term used to describe chronic inflammation found in the lining of the intestinal tract. IBD encompasses two major diseases—Crohn's disease (CD) and ulcerative colitis (UC) with additional distinctions based on the anatomical extent of disease and affiliated systemic clinical symptoms.[1–6] Although CD can affect any part of the gastrointestinal tract, it is most often found at the end of the small intestine and the beginning of the large intestine. UC, on the other hand, is confined to the large intestine (colon) and the rectum. UC begins in the rectum and extends contiguously up the colon. Neutrophils are predominant in acute UC, but lymphocytes, plasma cells, and eosinophils without granulomas are present in chronic disease. In contrast,

Departments of [1]Neurology, and [2]Biochemistry, Molecular and Cellular Biology, Georgetown University Medical Center, Washington, District of Columbia, USA.
[3]Dahlgren Memorial Library, Graduate Health and Life Sciences Research Library, Georgetown University Medical Center, Washington, District of Columbia, USA.
[4]Division of Rheumatology, Immunology and Allergy, Department of Medicine, Georgetown University Medical Center, Washington, District of Columbia, USA.

*Address correspondence to: Sona Vasudevan, PhD, Department of Biochemistry, Molecular and Cellular Biology, Georgetown University Medical Center, 3900 Reservoir Road, NW, Washington, DC 20057, USA, E-mail: sv67@georgetown.edu

the transmural granulomatous inflammation of CD is most frequently found in the distal small intestine and proximal colon but its "skip lesions" can affect any part of the gastrointestinal tract from the distal colon to the mouth.[7]

CD and UC are multifactorial diseases, with complex etiologies and risk factors. The incidence of each disease is about 3 to 19 cases per 100,000 persons per year with onset between 15 and 40 years. Smoking is a risk factor for aggressive CD, but it may be relatively protective for UC.[8] Other risk factors include urban living, appendectomy, tonsillectomy, and vitamin D deficiency.[9]

CD has immune deviation toward T helper type 1 (Th1) lymphocytes with elevated interferon γ (*IFN-γ*) and T-bet cells whereas UC is shifted toward the T helper type 2 (Th2) pathway with increased interleukin 13 (*IL-13*) and *GATA3* expression.[10] However, T helper type 17 (Th17) lymphocytes that express IL-17 and have *RORγ*t as their pivotal transcription factor, and regulatory T cells (Treg:*FOXP3*) are also dysregulated in these diseases. Innate immune cells such as macrophages, granulocytes, and dendritic cells participate by secreting *IL23*, which promotes proliferation of Th17 cells, innate lymphoid cells type 3, granulocytes, and natural killer cells.[11] A role for *IL23* in CD and UC is indicated by the success of monoclonal antibody therapies.[12,13]

Cellular immune mechanisms injure enterocytes; which disrupts epithelial barrier function[14] and alters interactions with the gut microbiome.[4,15–17] and new diagnostic methods have not evolved to take advantage of these alterations. The advent of high-throughput sequencing technologies has allowed for the leveraging of *in silico* research methods such as Genome Wide Association Studies (GWAS), which have identified more than 240 IBD associated genetic loci.[18,19] Only about 10–20% of GWAS loci are located in protein coding regions, whereas 80–90% are in noncoding regions and so exert their pathogenic effects by modulating gene expression.[20]

A problem affecting genetic and other aspects of IBD research has been the tendency to consider CD and UC as equivalent pathological processes. The search for genes across a wide spectrum of inflammatory gastrointestinal cases may fail to distinguish between genotypes selective for UC, CD, other systemic immune syndromes, and those for the IBD-unclassified category that is defined by exclusion as being neither UC nor CD. The importance of disease diagnosis is exemplified by twin studies that have revealed a proband concordance rate in monozygotic twins of around 58–62% for CD but 6–18% for UC that indicate a stronger genetic diathesis in CD than UC, and with major environmental influences on disease pathogenesis.[21–24] Another important confounder is the overlap with other autoimmune inflammatory diseases that may share common mechanisms of immune cell histopathology directed at specific ligands that may be widely distributed across the body in diverse target organs.

Treatment to date has focused on alleviating symptoms using nonspecific anti-inflammatory drugs such as sulfa derivatives, corticosteroids, Janus kinase inhibitors, and monoclonal antibodies directed predominantly at cytokines.[25–32] These potent immunomodulators can impair systemic immune responses and are associated with their own sets of adverse events linked to inhibition of their therapeutic targets. An aim of this work is to better define pathways of disease and to screen for more precise and specific targets directed at CD or UC specific pathological mechanisms. This requires an integrated approach that utilizes genomics, proteomics, human single-cell transcriptomes from biopsies, and other "big data" resources to build an accurate platform of IBD phenotypes.

We use a Systems and Network Medicine view to explore these diseases. Network medicine is based on the hypothesis that the impact of a single mutated gene is propagated along the gene's network links. These links can be envisioned as defining a disease module. Identification and characterization of these disease modules will help shed some light on the differences and commonalities that exist between these disease modules within their biological pathways; and they may reveal common or different biological processes (BPs) and functions that may exist between them.[33,34]

In the present study, we will consider genes that are not classified into CD or UC as IBD-unclassified. Although there are a number of common genes shared between the diseases (IBD-unclassified, CD and UC), the goal of this article is to focus on the differences between what defines UC and CD specifically to identify pathways that may distinguish them, and thus point to newer treatment modalities and targets.

In this study using the CoVex[35] systems medicine and drug-repurposing platform, we investigate IBD-unclassified, CD and UC as independent disease modules through an unbiased human protein–protein interaction (PPI) network and drug–target interactions. Given a list of user-defined disease genes (referred to as seeds) as input *via* the platform's custom proteins feature, we search the human interactome for viable drug–targets to identify repurposable drug candidates. We used the Multi-Steiner tree (MuST) algorithm implemented within the platform to expand each disease module by incorporating candidate genes with significant connections to the disease-related seed genes. Such expanded disease modules will identify a larger set of potential targets and drugs. We used the Closeness Centrality algorithm implemented within CoVex to search for newer targets and repurposable drugs. Through a network medicine approach, we provide a mechanistic view of the diseases and point to newer drugs and targets.

## Materials and Methods
### Disease genes (seeds)
An initial set of genes implicated in UC, CD, and IBD-unclassified were mined from literature by using the text-mining tool Linguimatics and direct searches using PubMed (https://www.ncbi.nlm.nih.gov/pubmed). The obtained genes were then checked against the curated dataset available through DisGeNET[36] and Uniprot.[37] Only genes that were curated in DisGeNET and Uniprot were retained for further analysis. This eliminates the bias that exists in the literature curation process. The set of genes thus obtained will be referred to as seeds.

### Drug-repurposing and generation of expanded disease module using CoVex platform
Recently, an interactive online platform (CoVex) was developed[35] as a one-stop shop systems medicine platform enabling network-based drug-repurposing. CoVex's systems medicine platform incorporates PPIs and drug (target) identification algorithms. Although the platform was developed for the severe acute respiratory syndrome coronavirus-2 virus, the algorithms and tools are applicable to any disease *via* the Custom Proteins feature implemented in the platform. Using the MuST algorithm implemented within CoVex, the initial set of seeds was expanded to include more protein–protein interacting partners with the seeds. These proteins will be referred to as the MuST proteins/genes. Twenty trees were generated with a hub-penalty of 0. The Closeness Centrality algorithm implemented within the platform was used for extracting drug targets and for ranking drugs.[38] The top 50 ranked drugs were extracted. We included only approved drugs and those that interacted directly with the proteins in the network. The main goal is to identify newer drug targets and potentially prioritize therapeutics for CD and UC. The output from CoVex in the GraphML format was input into Cytoscape for network visualization.

### Variant/single-nucleotide polymorphism information
The GWAS Catalog was used to mine for statistically significant single-nucleotide polymorphisms (SNPs) with a *p*-value of <0.05. The catalog was downloaded from https://www.ebi.ac.uk/gwas/downloads on February 4, 2021.[39,40] Genes implicated in IBD-unclassified, CD, and UC were obtained by filtering the Disease/Trait column of the catalog and searching for IBD, UC, and CD individually. The term IBD-unclassified does not exist in the GWAS catalog. Hence we used IBD as the search term. The genes that were unique to IBD were taken as IBD-unclassified. Since the goal of GWAS is to identify causative-SNPs in the genes, we included only the genes where a missense variant was identified. We have not included the genes that carried SNPs in the intronic, intergenic, and the regulatory regions because limited mechanistic information is available on their roles in disease progression.

### Protein–Protein interactions
In addition to the Integrated Protein–Protein Interactions database implemented in CoVex, we also used the Human Integrated Protein–Protein Interaction rEference (HIPPIE) database as a cross-referencing resource.[41] HIPPIE is a curated, publicly, and freely available resource for mining PPI.[41] The resource provides scored PPIs. HIPPIE integrates data from several other curated PPI resources. HIPPIE is very reliable, as it provides confidence scores based on experimentation. HIPPIE's confidence score ranges from 0 to 1 and reflects the quality of experimental evidence supporting each PPI. The HIPPIE data are based on confidence levels defined as quartiles of all confidence scores: medium confidence (0.63—second quartile of the HIPPIE score distribution) or high confidence (0.73—third quartile). To balance network coverage and reasonable PPI confidence, we used all PPIs that satisfy a medium confidence score of 0.63. The most recent update released on February 14, 2019 includes approximately over 295,000 experimentally determined PPIs between 17,000 human proteins.[42] The HIPPIE (v2.2) dataset was pre-filtered to remove interactions from non-human sources and those below the predefined medium-confidence score threshold of 0.63 to create a more stringent and reliable interactome.

### Semantic similarities between gene ontology terms

Semantic similarities between gene ontology (GO) terms [GO Biological Process (BP); Go Cellular component (CC); Go Molecular Function (MF)] for the MuST genes were calculated by using the GOGO software.[43] GOGO generates 2 output files. The first file contains the semantic similarity between pairs of genes, and the second file contains the gene functional clusters produced using an affinity propagation algorithm.[43] The ClueGO app implemented within the Cytoscape tool[44] was used for visualizing the results of GOGO to identify the enriched GO terms of the genes in the expanded disease module.

### Identification of relevant subnetwork within a disease network

Identification of a set of connected nodes within a huge network that is biologically relevant and meaningful is very important in Network Medicine. This is especially relevant in bigger networks that search for closely connected nodes when searching for drug targets. The diffusion app implemented within the Cytoscape tool,[44] based on a network propagation algorithm, was used to generate subnetworks.[45]

### Visualization

The widely used Cytoscape tool[44] was used for visualizing various networks. The ClueGo plug-in app implemented within Cytoscape was used to decipher functionally grouped gene ontologies.[46] The MyVenn tool implemented in the Comparative Toxicogenomics Database was used for creating the Venn diagrams.[47]

### Pathway analysis

Several tools were independently used to map the genes to pathways since we do not have one consolidated pathway tool that provides unambiguous results. The seed proteins and the proteins from the expanded module were mapped to Kyoto Encyclopedia of Genes and Genomes (KEGG) pathways,[48] Reactome pathway knowledgebase,[49] WikiPathways (WP),[50] Ingenuity Pathway Analysis tool (IPA), and the Pathway Studio tool.[51] The KEGG, Reactome, and WP pathway tools were assessed through the gprofiler server.[52]

### Gene functional classification

The Database for Annotation, Visualization and Integrated Discovery (DAVID) tool was used for gene functional analysis.[53]

## Results and Discussion

### Disease seed proteins

The initial set of seed proteins obtained from the literature and confirmed by curated resources DisgeNet and Uniprot is provided in Supplementary Table S1. Figure 1A shows the seed genes that is common and unique between CD, UC, and IBD-unclassified. Figure 1B shows the genes unique to CD and UC and those that are common between them. Disease module construction requires an initial set of disease genes and a molecular network. The HIPPIE data resource was used to create a PPI. Enough evidence exists in the literature that disease-associated proteins cluster together in a network neighborhood through physical interactions in an interactome.[54–56] CD, UC, and IBD-unclassified follow this expectation of its genes connected as evaluated using the HIPPIE PPI database.

### Generation of disease modules from seed proteins using the MuST algorithm

The next step was to use the connected seed proteins to generate an expanded disease module using the MuST algorithm implemented within the CoVex platform. Hereafter, these will be referred to as MuST proteins. The MuST algorithm as mentioned earlier has been used for the identification of topological neighborhoods of seed proteins based on the significance of their connections to the seed proteins. Figure 2 shows the total number of MuST proteins thus obtained starting from the seed proteins. To make sure that the expanded proteins made biological sense, they were evaluated for their biological evidence using the GO terms and sematic similarities using the GOGO software. The GOGO software produces semantic similarities between gene pairs and clustering of the genes based on their functional similarities. A similarity score > 0.5 is typically considered significant, but we increased that threshold to 0.6. The MuST algorithm expanded the number of proteins from 31 unique seed proteins to 104 for CD, 42 to 149 for UC, and 17 to 81 for IBD-unclassified.

### Topological relationship between CD, UC and IBD-unclassified

CD and UC, while having features of being distinct diseases, share common mechanisms and pathogenies as both fall under the umbrella of IBD. There does exist a significant relationship between activated genes and molecular events that lead to inflammatory diseases.

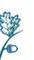

**FIG. 1.** **(A)** Venn diagram showing the total number of starting seed proteins. There are total of 45 in CD (labeled CD_Seeds), 42 in UC (labeled UC_Seeds), and 30 in IBD-unclassified (labeled IBD_Seeds). The Venn diagram was created by using the MyVenn tool in the CTD database. **(B)** Seed genes that are common between CD and UC are shown in green. The unique genes within CD and UC are shown as dark blue squares. The light blue circles at the center of the network represents Ulcerative Colitis and Crohn's disease. CD, Crohn's disease; CTD, Comparative Toxicogenomics Database; IBD, inflammatory bowel disease; UC, ulcerative colitis.

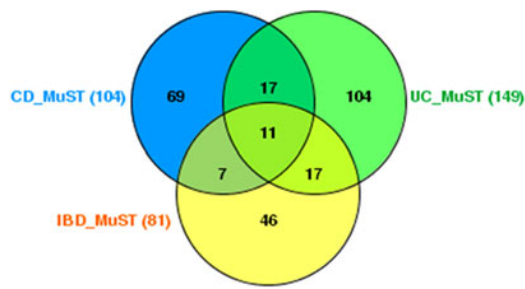

**FIG. 2.** Venn diagram showing the total number of expanded MuST proteins from the seed proteins. There are a total of 104 in CD (labeled CD_MuST), 149 in UC (labeled UC_MuST), and 81 in IBD-unclassified (labeled IBD_MuST). The Venn diagram was created using the MyVenn tool in the CTD database.

The results of the semantic similarity and clustering analysis for IBD-unclassified, CD, and UC are provided in Supplementary Table S2 (Sheets 1–3) that includes GO BP terms with a >0.6 cutoff. We used the results of Gene functional similarity analysis produced by the GOGO software to visualize the distinct BPs between CD and UC using the ClueGO Cytoscape plug-in app. Out of a total of 105 GO BP terms for CD and 95 for UC, 50 BP terms were distinct for CD and 40 for UC, again pointing to the fact that they operate by different underlying biological mechanisms. The corresponding results are provided in Figure 3.

### Pathway analysis between CD, UC, and IBD-unclassified

We used several pathway analysis tools to map the MuST proteins to see both the common mechanisms and distinctive pathways that distinguish them in the quest to find newer targets and drugs for them. The results of the top 10 or so statistically significant pathways from each of the pathway tools KEGG, Reactome, WP, IPA, and Pathway Studio are provided in Supplementary Table S3. The gprofiler web server was used for functional enrichment analysis and to extract the pathways for KEGG, Reactome and WP. Although each tool is based on a rigorous methodology to pull up the significant pathways, not surprisingly, we did not find any pathways that were common among the various pathway analysis tools. This is an issue generally in the field currently, and there is no consensus among these various tools. Each of the tools is developed based on different resources and hence the results can be considered to be more complementary. Combining information from all of the tools is useful for global biological interpretations. We have used IPA and Pathway Studio in our discussions, as it suggests more possible links between differentially regulated genes and with genes in highly relevant networks. The outcomes were different but provide distinctly different perspectives on potential pathological mechanisms.

DAVID Gene Functional Classification tool clustered the CD genes into five categories (Table 1). Nuclear and lipid receptors included *ESR1*, *PPARA*, *PPARG*, *NR1H4*, and *VDR*. *TLR4*, *TLR8*, *MYD88*, and *NLRP3* are involved in Pathogen-Associated Molecular Patterns. Chaperones, *TGFBR1* and *IL2RA* (*CD25*) were in the other groups. IPA pathways, upstream regulators, and causal networks supported cytokine activation of *STAT5B* with upregulation of prion inflammatory cytokines (Table 2). *ESR1*–*FOXO1* interactions are involved in obesity, which is a risk factor for CD.[57,58]

*PPARG*, *PPARA*, *NR1H3*, and *AIP* are lipid receptors that are complex with retinoid × receptor to regulate histone deacetylation, suppress transcription, and induce genes for fatty acid oxidation. *BATF* is activated by *STAT3* and enhances Th17, T follicular helper and CD8+ dendritic cell production, and B cell heavy chain switching. *CEBPB* is involved in Th2 lymphocyte differentiation, whereas a variant is associated with juvenile polyposis. Currently approved kinase inhibitors used for IBD interact with *TGFBR1*, *MAP2K1*, *PRKAA2*, and *TYK2*.[59]

Pathway Studio provided an alternative perspective. The BPs were organized around *STAT* mechanisms involved in signaling from *IL1B*, *TNF*, *TGFB*, and retinoic acid. *Leptin*, *CCL2*, *FOXO1*, *PPARG*, *TLR4*, and *MYD88* were implicated in the adipokines production by the adipocyte pathway. Treg function was suggested by *TGFBR1*, *IL2RA*, *SMAD3*, *STAT5A*, *PTPN6*, and *MAP2K1* (Treg-Cell Differentiation) and *TYK2* in Th17 cells (*AHR* Signaling in Th17 Cells Function). Xenobiotics and environmental endocrine disrupters are implicated by *AIP* in the MuST gene list.

The UC gene list, on the other hand, was clustered in DAVID into five groups: proteasome subunits, serine-threonine kinases, transcription factors, G-protein coupled receptors for cytokines, and receptors that activate

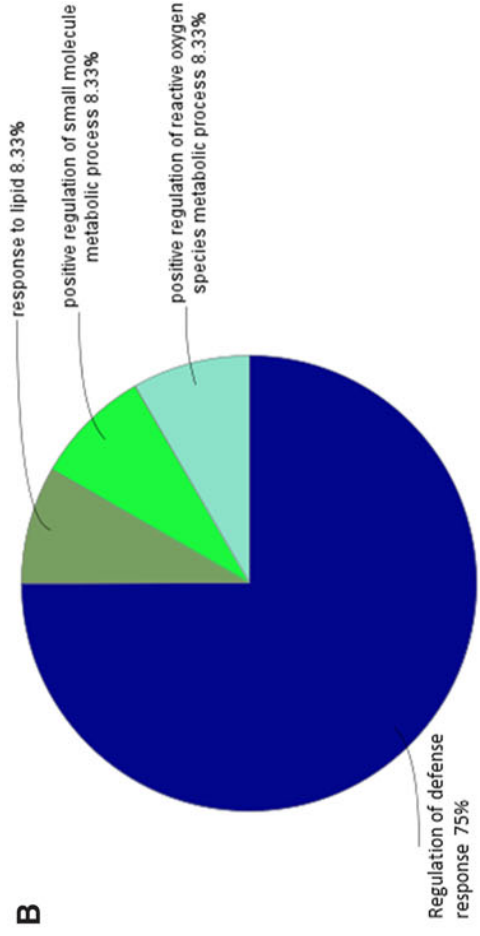
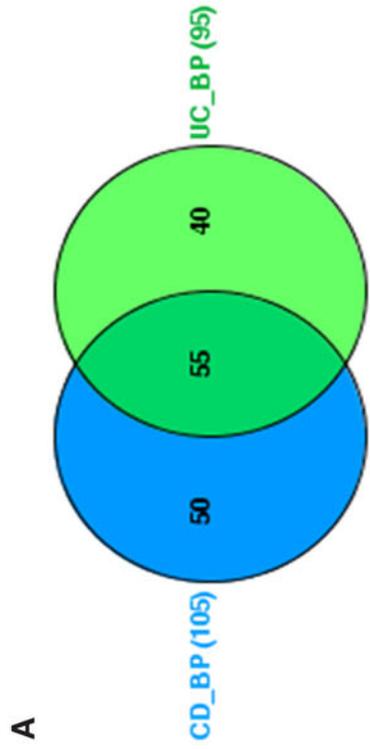
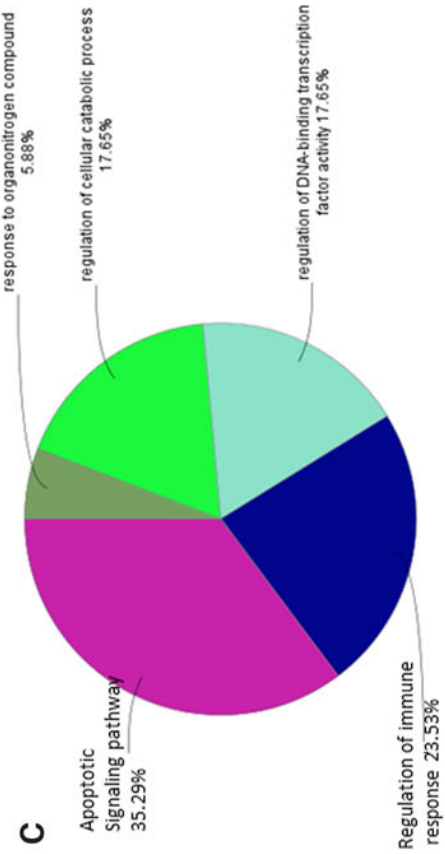

**FIG. 3.** Results of GO BP analysis using ClueGo cytoscape Plug in. **(A)** Venn diagram showing the common BP Go ontology terms and distinct BP terms between CD and UC. MyVenn tool available through the CTD database was used. **(B)** The distinct BP terms for CD. **(C)** The distinct BP terms for UC. GO, gene ontology; BP, biological process.

**Table 1. Gene Functional Analysis Using the Database for Annotation, Visualization and Integrated Discovery Tool**

| IBD | CD | UC |
|---|---|---|
| Gene group 1 Enrichment score: 3.2 EIF3E EEF1G SFN | Gene group 1 Enrichment score: 4.1 ESR1 FOXO1 PPARA PPARG VDR NR1H4 TRIM28 NCOA2 RUNX1 NCOA1 ZBTB16 ZMIZ1 | Gene group 1 Enrichment score: 3.2 PSMA4 PSMA1 PSMB4 PSMB2 |
| Gene group 2 Enrichment score: 2.2 ERBB2 GSK3B BUB1 CASK | Gene group 2 Enrichment score: 2.9 ENO1 SF3A1 SRSF1 HSPA8 HNRNPK | Gene group 2 Enrichment score: 2.4 MAPKAPK2 CDK15 STK11 STK4 |
| Gene group 3 Enrichment score: 1.9 SFRP1 SFRP2 WNT4 | Gene group 3 Enrichment score: 1.9 TLR4 TLR9 MYD88 NLRP3 NLR | Gene group 3 Enrichment score: 2.3 SNAI1 IKZF1 IKAROS PRDM1 ZNF281 |
| Gene group 4 Enrichment score: 1.7 PTPRK SLAMF8 SLAM ITGAL ITGA4 ITGB8 EPOR | Gene group 4 Enrichment score: 1.8 TGFBR1 PRKAA2 MAP2K1 | Gene group 4 Enrichment score: 1.7 CXCR2 CCR6 CXCR1 TBXA2R FCGR2A |
| | Gene group 5 Enrichment score: 1.8 IL2RA FPR2 ITGAM ERAP1 ERAP2 FUT2 NCLN | Gene group 5 Enrichment score: 1.2 IL1R2 IL1RAP FCGR2A IL7RPTPRD ICOSLG VCAM1 |

CD, Crohn's disease; IBD, inflammatory bowel disease; UC, ulcerative colitis.

**Table 2. Top 5 Statistically Enriched Canonical Pathways, Upstream Regulators, Causal Networks for Crohn's Disease, Ulcerative Colitis, and Inflammatory Bowel Disease-Unclassified Identified Using Ingenuity Pathway Studio**

| Pathways | Upstream regulators | Causal network |
|---|---|---|
| **CD** | | |
| • Glucocorticoid receptor signaling (n = 16) | CEBPB | BATF |
| • Hepatic fibrosis/hepatic stellate cell activation (n = 9) | SP1 RELA | AIP NR1H3 (9-cis-retinoic acid receptor) |
| • Role of hypercytokinemia/ hyperchemokinemia in the pathogenesis of influenza (n = 7) | FOXO1 PPARG | PPARA (NR1C1) STAT5B |
| • Hepatic fibrosis signaling pathway (n = 11) | | |
| • Osteoarthritis pathway (n = 9) | | |
| **UC** | | |
| • Cardiac hypertrophy signaling (enhanced; n = 19) | TP53 | RORC |
| • Osteoarthritis pathway (n = 12) | NR3C1 | FHL1 |
| • PI3K/AKT signaling (n = 11) | NFKB1 | EGLN2 |
| • IL-6 signaling (n = 9) | JUN | SRF |
| • Granulocyte adhesion and diapedesis (n = 10) | FOS | NFKB1 |
| **IBD** | | |
| • Protein kinase A signaling (n = 10) | THRB | PIM2 |
| • Role of osteoblasts and osteoclasts Chondrocytes in rheumatoid arthritis (n = 8) | TP53 MYC | SMARCB1 UBD |
| • Axonal guidance signaling (n = 10) | PAX2 E2F3 | |
| • Wnt/-catenin signaling (n = 7) | MAP3K8 | |
| • Molecular mechanisms of cancer (n = 9) | | |

The numbers of proteins in each pathway are given in parenthesis.

JAK-STAT pathways or were decoys that inactivate IL1B. Ligands for these receptors include IL8, MIP3A, TBXA2R, IL7, IgG, and ITGA4/ITGB1. The UC pathways were related to neutrophils (CXCL2, CXCL8, CXCR1, IL1B, VCAM1, TICAM1), lactoferrin, C3 complement factor[60] FCGR2A, and endogenous peptide antigen presentation involving proteasome components.

The IPA pathways were related to granulocyte function, IL6, and PI3K–AKT signaling. Upstream factors suggested interactions between the proinflammatory transcription factors NFKB, Jun, and Fos and anti-inflammatory/transcription system suppressors TP53 and NR3C1 (glucocorticoid receptor). Downstream effectors included EGLN2 that can activate NFKB, and the general transcription activation SRF. Oddly, RORC, the canonical transcription factor for Th17 cells, was generated. These pathways had modest overlap with those cited earlier.[61–63]

Neutrophil activation was supported by IL1B and IL8 (CXCL8) and its receptors (CXCR1 and CXCR2) and neutrophil granule proteins MPO, MMP9, and LTF. MMP9 is required for IL1 activation. IL1B is inactivated by binding its decoy receptor IL1R2 and IL1RAP. Complement activation was supported by C3 and MASP2. Apoptosis was suggested by involvement of RIPK2, CASP3, CASP4, YWHAE, and BIRC3. Macrophage activation was suggested from CCL20 (MIP3A) and its receptor, CCR6, and production of IL6. Cell surface interactions were mediated by VICAM, TICAM, and LGALS1. Receptor signaling induced cascades of cAMP (ADCY7), tyrosine kinases (FYN, PIK3R1, PTPRD), serine-threonine kinases (MTOR), and MAP kinase. Protein catabolization was inferred by proteasome subunit expression. Extensive gene regulation was inferred by histone deacetylases, transcription factors, and related proteins. It is possible that these participated in chromatin condensation to form neutrophil extracellular traps in the NETosis pathway of regulated neutrophil death. A role for thromboxane A2 was inferred by the presence of its receptor, TBXA2R; this may be a novel target for drug development.

### Drug-repurposing

One of the main goals of this study was to identify newer drug targets and approved drugs that can be re-purposed for UC and CD. We used the Closeness Centrality algorithm implemented in the CoVex platform for extracting the drugs that directly interacted with our MuST proteins. The top 50 approved drugs that mapped to their target genes were extracted. Drugs mapped directly to the target genes generated by the application of Closeness Centrality algorithm is provided in Table 3. The ranking of the drugs is provided in Supplementary Table S4. We used the Diffusion Cytoscape app to generate subnetworks of nodes that had many edge mappings to drugs for UC and CD (Figs. 4 and 5).

*FYN*, a non-receptor tyrosine protein kinase that plays a role in many BPs that include regulation of cell growth and survival, cell adhesion, and integrin-mediated signaling, directly connected with about 31 drugs. The majority of the drugs are kinase inhibitors. Currently, according to opentargets.org,[64] there are a total of 418 clinical trials in Phase IV, 300 in Phase III, 133 in Phase II, and 65 in Phase I with 91 associated targets. Twenty of the genes that are currently targets are in our MuST expanded UC gene set. Interestingly, the drug that is Ranked 29, Tofacitinib is also currently in Clinical Trials. *JAK2*, another non-receptor Tyrosine protein kinase, which is currently a target in clinical trials, shares some of the kinase inhibitors that directly interact with *FYN*.[65–67] *FYN* and *JAK2* are functionally very similar. *JAK2* mediates essential signaling events in both innate and adaptive immunity. In the cytoplasm, it associates with both type 1 and type 2 receptors, which include *IFN-β* and *IFN-γ* and several ILs. Other genes that potentially need to be explored further are STK4, a serine-threonine protein kinase, and NTRK1, a high-affinity nerve growth factor.

On the other hand, for CD the rich nodes were *ESR1*, nuclear hormone receptor, *SMAD3*, an intracellular signal transducer and transcriptional modulator, and *NR3C1*, Glucocorticoireceptor, *TYK2*, a non-receptor tyrosine kinase involved in IFN-signaling, were among the proteins that contained the maximum nodes. Interestingly, *NR3C1* is currently a target in many ongoing clinical trials (opentargets.org). Our analysis points to several other potential targets such as *ESR1* and others. Interestingly, several kinase inhibitors were among the top ranked drugs such as Nintedenib and Fostanaib as was seen in UC. However, the targets of these drugs are different, implying that they work differently in UC and CD. So, potentially the same drug can be given for both these diseases, which could potentially be mechanistically different.

### Mapping of curated variants identified from GWAS studies

Supplementary Table S1 (Sheet 2) provides the curated list of missense variants identified from GWAS studies. Interestingly, we find SNPs in several genes that mapped to our expanded disease module (*TNFSF15, NOD2, CARD9, FCGR2A, IL23R, MST1, PLCG2, ADCY7*). These affect the various pathways discussed earlier; they will be explored in the future and are beyond the scope of this present study.

**Table 3. Drugs Obtained by the Application of the Closeness Centrality Algorithm**

| CD | UC | IBD |
|---|---|---|
| Fostamatinib | Fostamatinib | Fostamatinib |
| Nintedanib | Bosutinib | Bosutinib |
| Zinc acetate | Sunitinib | Sunitinib |
| Zinc chloride | Nintedanib | Nintedanib |
| Zinc | Zinc acetate | Midostaurin |
| Bosutinib | Zinc chloride | Crizotinib |
| Sunitinib | Zinc | Dasatinib |
| Copper | Midostaurin | Ruxolitinib |
| Midostaurin | Crizotinib | Erlotinib |
| Ruxolitinib | Copper | Sorafenib |
| Dasatinib | Ruxolitinib | Vandetanib |
| Crizotinib | Sorafenib | Axitinib |
| Axitinib | Dasatinib | Copper |
| Vandetanib | Axitinib | Neratinib |
| Neratinib | Vandetanib | Zinc acetate |
| Hexachlorophene | Neratinib | Zinc chloride |
| Imatinib | Pazopanib | Zinc |
| Tamoxifen | Erlotinib | Gefitinib |
| Erlotinib | Imatinib | Tamoxifen |
| Pazopanib | Gefitinib | Imatinib |
| Resveratrol | Hexachlorophene | Pazopanib |
| Arsenic trioxide | Ceritinib | Hexachlorophene |
| Acetylsalicylic acid | Tamoxifen | Acetylsalicylic acid |
| Diethylstilbestrol | Niclosamide | Ceritinib |
| Niclosamide | Acetylsalicylic acid | Niclosamide |
| Estradiol | Afatinib | Resveratrol |
| Clotrimazole | Resveratrol | Afatinib |
| Mifepristone | Nilotinib | Lapatinib |
| Bithionol | Arsenic trioxide | Astemizole |
| Foreskin keratinocyte (neonatal) | Clotrimazole | Nilotinib |
| Cisplatin | Bortezomib | Bithionol |
| Clomifene | Fluphenazine | Clotrimazole |
| Pseudoephedrine | | Clomifene |
| Dobutamine | | Paclitaxel |
| Raloxifene | | Doxorubicin |
| Sulfasalazine | | Fluphenazine |
| Danazol | | Tofacitinib |
| Podofilox | | Docetaxel |
| Progesterone | | Miconazole |
| Ethinylestradiol | | Chlorpromazine |
| | | Diethylstilbestrol |
| | | Regorafenib |
| | | Adenosine |
| | | Podofilox |
| | | Estradiol |
| | | Diiodohydroxyquinoline |

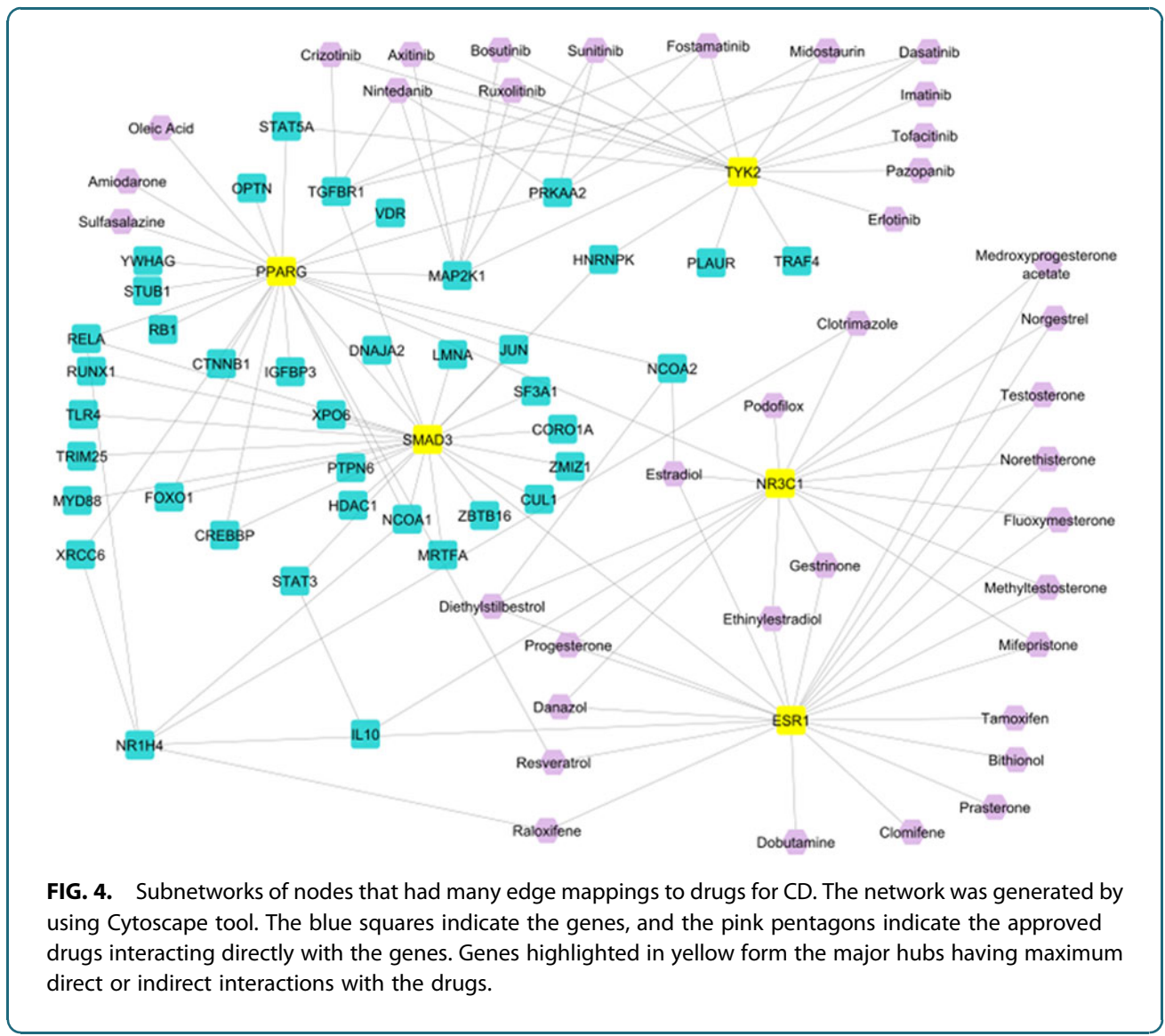

**FIG. 4.** Subnetworks of nodes that had many edge mappings to drugs for CD. The network was generated by using Cytoscape tool. The blue squares indicate the genes, and the pink pentagons indicate the approved drugs interacting directly with the genes. Genes highlighted in yellow form the major hubs having maximum direct or indirect interactions with the drugs.

**Conclusion**

We have provided a comprehensive Systems and Network medicine approach to the study IBD. CD, UC, and IBD-unclassified share many genes, pathways, and processes that bring them together. In the clinic, there are many cases that are still difficult to be classified as CD or UC and are currently labeled as Intermediate Colitis. We need to extensively study these cases and see whether they overlap with our IBD-unclassified gene set. The question arises as to whether a third distinct disease can be defined.

This approach can be applied, in principle, to any disease. Our limitations in this study design are the lack of gene-specific granularities that exist within the subtypes of CD, UC, and IBD-unclassified. According to the Montreal Classification,[68] UC is sub classified on the basis of the extent of the disease: proctitis-limited to the rectum; proctosigmoiditis and sigmoid colon; left-sided colitis—up to the splenic flexure; or extensive colitis-proximal to the splenic flexure. The CD is classified based on its location into ileal, colonic, ileocolonic, and isolated upper disease, with ileocolonic being the most common form of CD.

We have mentioned earlier and a well-known fact is that CD has a trend for immune deviation toward Th1 lymphocytes and UC is shifted toward the Th2 pathway. This binary polarization is too simplistic, because other immune cell types and mechanisms are also dysfunctional, including Th17 lymphocytes that express *IL-17*, *IL21*, and *IL22*, and have *RORC* as their pivotal transcription factor, and Treg that secrete *TGFB* and *IL10* under control of *FOXP3*. Innate immune cells such as macrophages, granulocytes, and dendritic cells play an integral role by secreting *IL23*, which promotes proliferation of Th17, innate lymphoid type 3, granulocytes, and natural killer cells.[11] Tracing the source(s) of individual cytokines such as *TNF*, *IL6*, and *IFN-γ* from these diverse cell populations is problematic and will require single-cell transcriptome studies for clarification.[69]

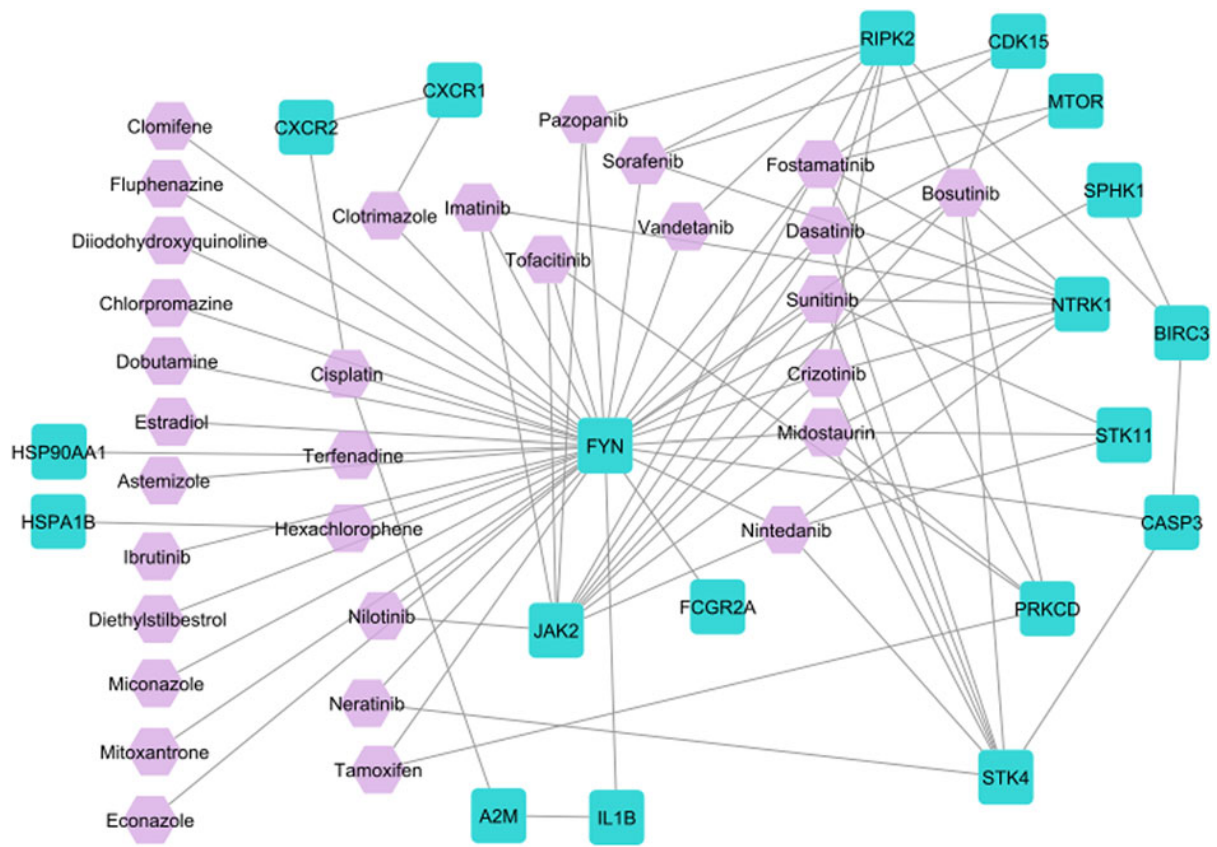

**FIG. 5.** Subnetworks of nodes that had many edge mappings to drugs for UC. The network was generated by using Cytoscape tool. The blue squares indicate the genes, and the pink pentagons indicate the approved drugs interacting directly with the genes.

Recently, the creation of a Gut Atlas is underway. This will uncover the distinctions between these diseases more granularly. Single-cell RNAseq data are, indeed, accumulating.[70,71] Interestingly, some of the genes uncovered from these studies are present in our datasets. Longitudinal biopsy studies with single-cell sequencing may become the standard for diagnosis, grading severity, selecting appropriate mechanistically targeted therapies, and as an assessment tool in clinical trials. In addition, we need to construct Diseasomes that will shed more light on the co-morbidities that exist with CD, UC, and IBD-unclassified and the crucial role of the microbiome.

One of our long-term goals is to establish a clearing house of data from curated databases with user-friendly software and workflows to facilitate the integration of big data analytics for informatics studies into clinical practice. The complexity of machine-learning and statistical outcomes may be overwhelming for the practicing clinician and so it will be important to develop adaptable interfaces that facilitate information flow for clinical decision making. The tools are not transparent and there is a need for clinical validation of informatics outcomes.

## Authors' Contributions
S.V. conceived the idea, validated the data, drove the analyses and discussions, and wrote the article. T.K. contributed to several important discussions of the article. S.C.M. helped with the Network analysis. R.M. developed several scripts for the analysis. C.G. carried out curation of the literature data. Jonathan Hartmann provided text mining data by using the Linguamatics tool. J.B. contributed to writing part of the article and participated in extensive discussions.

## Acknowledgments
The authors thank Ritika Kundra, who, as a graduate student at Georgetown University, paved the way for this work. They thank Sarma Dittakavi, Professor Emeritus, Laboratory Medicine and Pathobiology, University of Toronto, and Rajaram Gana, Georgetown University, for the many stimulating discussions that they had together. They are grateful to Dr. Vasantha Kolachala, Emory School of Medicine, for providing insights into single-cell RNAseq analysis. They thank Elliott Crooke, Professor and Chair, Department of Biochemistry and Molecular and Cellular Biology, and Senior Associate Dean, Faculty and Academic Affairs, Georgetown University, for his support and encouragement. This work is dedicated to the author's sister, the late Shanthi Sitaraman, who was a physician-scientist at the Emory School of Medicine.

## Disclaimer
This article has been submitted solely to this journal and is not published, in press, or submitted elsewhere, for peer review. The article has, however, been uploaded to the well-known preprint server: arXiv.org. Subjects: Molecular Networks (q-bio.MN) Cited as: arXiv:2006.04181.

## Ethics Approval and Consent to Participate
Not applicable. This study does not utilize any data that require ethical approval or consent to participate.

## Author Approval
All authors had access to the study data and had reviewed and approved the final article.

## Author Disclosure Statement
The authors declare that they have no competing interests.

## Funding Information
No funding was received for this work.

## Supplementary Material
Supplementary Table S1  
Supplementary Table S2  
Supplementary Table S3  
Supplementary Table S4

## Abbreviations Used

- AIP = aryl hydrocarbon interacting protein
- BP = biological process
- CD = Crohn's disease
- CTD = Comparative Toxicogenomics Database
- DAVID = Database for Annotation, Visualization and Integrated Discovery
- GO = gene ontology
- GWAS = Genome Wide Association Studies
- HIPPIE = Human Integrated Protein–Protein Interaction rEference
- IBD = inflammatory bowel disease
- IFN-$\gamma$ = interferon $\gamma$
- IL = interleukin
- ITGA4/ITGB1 = integrin $\alpha$-4/$\beta$-1
- MuST = Multi-Steiner tree
- PPI = protein–protein interaction
- Th1 = T helper type 1
- Th17 = T helper type 17
- Th2 = T helper type 2
- Treg = regulatory T cells
- UC = ulcerative colitis
- WP = WikiPathways